\documentclass[preprint,showpacs,preprintnumbers,amsmath,amssymb]{revtex4}

\usepackage[dvips]{graphics}
\usepackage{graphicx}
\usepackage{amsmath}
\usepackage{amssymb}

\begin{document}

\title[Kicked nonlinear quantum scissors  and entanglement generation]
{Kicked nonlinear quantum scissors  and entanglement generation}

\author{A. Kowalewska-Kud{\l}aszyk}
\affiliation{Faculty of Physics, Adam Mickiewicz
 University, Umultowska 85, 61-614 Pozna\'n, Poland}

\author{W. Leo\'nski}
\email{wleonski@if.uz.zgora.pl}
\author{T. Dung Nguyen}
\author{V. Cao Long}
\affiliation{Quantum Optics and Engineering Division, Institute of Physics, University of Zielona G\'ora, Z. Szafrana 4a, Zielona G\'ora 65-516, Poland}

\begin{abstract} 
We consider a nonlinear coupler with two Kerr-like oscillators mutually coupled by continuous linear interaction and excited by a series of ultra-short external pulses. We show that the system behaves as \textit{nonlinear quantum scissors}. It evolves such a way that it can be treated as \textit{qubit-qubit} system. We derive analytic formulas for the probabilities of the states involved in the system's evolution and show that they differ from those already discussed in the literature and corresponding to the continuously excited models. Moreover, for model discussed here, maximally entangled Bell states can be generated with high efficiency.
\end{abstract}


\maketitle

\section{Introduction}

Systems involving nonlinear or parametric oscillators were applied in numerous quantum optical models.  For instance, they were concerning generation of various quantum states of the field \cite{MTK90,TMK91,OP04,OAP040507}, quantum-optical properties of nonlinear structures \cite{SP10,P11,AGK11} Moreover, nonlinear oscillator models were considered in a context of Einstein-Podolsky-Rosen paradox \cite{AK06b}, construction various models of  \textit{quantum nonlinear scissors} (QNS) \cite{LT94,L9697,ML04,LM04,SWU0607,KL060910,KLP11,GSK12} or photon(phonon) blockade \cite{LMG10,MPL13}. In this paper we shall concentrate on a new model involving two quantum nonlinear oscillators that behaves as \textit{qubit-qubit} system and allow for generation of Bell states.

\section{The model and its solutions}

In this paper we discuss a model similar to that considered in \cite{LM04}, involving two nonlinear quantum oscillators that are characterized by Kerr-like nonlinearities $\chi_a$ and $\chi_b$, and labeled by $a$ and $b$ . The oscillators are mutually coupled by linear interaction and are excited by external electromagnetic field. In fact we deal here with Kerr-like nonlinear coupler discussed in numerous papers (for instance see \cite{KP97}) that is described by the following Hamiltonian expressed in terms of boson creation and annihilation operators $\hat{a}^\dagger$ ($\hat{b}^\dagger$) and $\hat{a}$ ($\hat{b}$), respectively:
\begin{equation}
\hat{H}_{NL}\,=\, \frac{\chi_a}{2}(\hat{a}^\dagger)^2\hat{a}^2+\frac{\chi_b}{2}(\hat{b}^\dagger)^2\hat{b}^2+
\epsilon\hat{a}^{\dagger}\hat{b}+\epsilon^*\hat{a}\hat{b}^{\dagger},
\label{eq1}
\end{equation} 
where $\epsilon$ describes the strength of internal coupler's coupling.
The system is externally excited in one mode and this excitation is in the form of series of ultra-short coherent pulses and differs in this point from the model discussed in \cite{LM04} where continuous excitation of constant amplitude was assumed. In particular we assume the interaction between the external classical field and the field of the quantum mode $a$ inside a coupler. This interaction can be modeled with use of \textit{Dirac-delta} function. In consequence, the Hamiltonian corresponding to this interaction can be written as:
\begin{equation}
\hat{H}_{K}=(\alpha\hat{a}^{\dag}+\alpha^*\hat{a})\,\sum\limits_{k=0}^{\infty}\delta(t-kT).
\label{eq2}
\end{equation}
The parameter $\alpha$ appearing here describes the strength of the external field -- nonlinear system interaction, $k$ enumerates external pulses, whereas $T$ is a time between two subsequent pulses. 

Since in this communication we restrict ourselves to the case of ideal situation, \textit{i.e.}, the model without damping processes, we shall describe the system's evolution in terms of the time-dependent wave function. It can be expressed in
the n-photon Fock basis as
\begin{equation}
|\Psi\rangle = \sum_{m,n=0}^{\infty} c_{m,n}|m\rangle_a |n\rangle_b,
\label{eq3}
\end{equation}
where $c_{m,n}$ are complex probability amplitudes, $|m\rangle_a$ and $|n\rangle_b$ are $n$-photon Fock states corresponding to the modes $a$ and $b$, respectively.

We assume that our system is externally pumped and the losses are neglected. Nevertheless, if we assume that the excitation is sufficiently weak, the system's dynamics will remain closed within the finite set of $n$-photon states. Thanks to the presence of the resonant coupling by zero-frequency component of external excitation between some eigenstates generated by the Hamiltonian $\hat{H}_{NL}$. For the system discussed here only four states are involved in the system's evolution. They are: $|0\rangle_a\otimes |0\rangle_b$, $|0\rangle_a\otimes |1\rangle_b$, $|1\rangle_a\otimes |0\rangle_b$ and $|1\rangle_a\otimes |1\rangle_b$. All these states correspond to the same eigenenergy of $\hat{H}_{NL}$ equal to zero. Hence, we can truncate the wave-function and it takes the following form:
\begin{equation}
|\Psi\rangle_{cut}\,=\,c_{0,0} |0\rangle_a |0\rangle_b + c_{0,1} |0\rangle_a |1\rangle_b + c_{1,0} |1\rangle_a |0 \rangle_b + c_{1,1} |1\rangle_a |1\rangle_b.
\label{eq4}
\end{equation}
Thus, using Schr\"odinger equation and applying standard procedure we can derive equations of motion determining probability amplitudes $c_{i,j}$, $\{i,j\}=\{0,1\}$. With use of the method shown in \cite{LDT97} we find solutions for the amplitudes corresponding to the moments of time just after $k$-th pulse. If we assume that for the time $t=0$ we have no photons in the system, \textit{i.e.} $|\Psi (t=0)\rangle=|0\rangle_a|0\rangle_b$ the amplitudes become:
\begin{eqnarray}
c_{0,0}(k)&=& \frac{1}{2\epsilon T\Omega}\left(
(2\alpha^2 - \Omega_2^2) \cos \frac{k\Omega_1}{\sqrt{2}} -
(2\alpha^2 - \Omega_1^2) \cos \frac{k\Omega_2}{\sqrt{2}}
\right)
 \nonumber\\
c_{0,1}(k)&=& \frac{\alpha}{\Omega}\left(
\cos \frac{k\Omega_1}{\sqrt{2}} - \cos \frac{k\Omega_2}{\sqrt{2}}
\right)
 \nonumber\\
c_{1,0}(k)&=& \frac{i\alpha}{\sqrt{2}\,\epsilon\,T\,\Omega\,\Omega_1\Omega_2}\left(
\left(\Omega_2^2 - 2(\epsilon^2T^2+\alpha^2)\right)\Omega_2\sin \frac{k\Omega_1}{\sqrt{2}} +
\epsilon T(\epsilon T-\Omega)\Omega_1\sin \frac{k\Omega_2}{\sqrt{2}}
\right)
 \nonumber\\
c_{1,1}(k)&=&\frac{i\sqrt{2}\alpha^2}{\Omega}\left(
\frac{1}{\Omega_2}\sin\frac{k\Omega_2}{\sqrt{2}}
-\frac{1}{\Omega_1}\sin\frac{k\Omega_1}{\sqrt{2}}
\right),
\label{eq5}
\end{eqnarray}
where the following frequencies were defined:
\begin{eqnarray}
\Omega\,=\,\sqrt{\epsilon^2T^2+4\alpha^2},   \nonumber \\
\Omega_1\,=\, \sqrt{\epsilon^2T^2+2\alpha^2+\epsilon\, T\,\Omega},   \quad
\Omega_2\,=\,\sqrt{\epsilon^2T^2+2\alpha^2-\epsilon\, T\,\Omega}.
\label{eq6}
\end{eqnarray}

This result is an extension of that discussed in \cite{LDT97}. If we assume here that there is no coupling between two modes ($\epsilon =0$) and the system's evolution starts form the state $|\Psi (t=0)\rangle=|0\rangle_a|0\rangle_b$ (we have no photons in both modes), the probability amplitudes $c_{0,1}=c_{1,1}=0$. Moreover, we have $c_{0,0}=\cos k\alpha$ and $c_{1,0}=-i\sin k\alpha$. In consequence, during the system's evolution, we have no photons in the mode $b$, whereas we can observe regular oscillations between the states $|0\rangle_a$ and $|1\rangle_a$. This result is identical to that discussed in \cite{LDT97}.

To check the validity of the solution and in consequence, exactness of the wave-function truncation we compare above analytical results with those of numerical calculations. Therefore, we define unitary evolution operators on a basis of the Hamiltonians (\ref{eq1}) and (\ref{eq2}). They are (we use units of $\hbar =1$):
\begin{equation}
\hat{U}_{NL}\,=\,\exp \left( -i \hat{H}_{NL}\,T \right)\quad\rm{and}\quad\hat{U}_K\,=\,\exp 
\left(
 -i (\alpha\hat{a}^{\dag}+\alpha^*\hat{a})
  \right),
\label{eq7}
\end{equation}
where the first of them ($\hat{U}_{NL}$) corresponds to the "free" evolution of the wave-function during the time between two subsequent pulses, whereas the latter ($\hat{U}_K$) describes the influence  of single infinitesimally short pulse. Thus, the product of these two operators transforms the wave function from that corresponding to the moment of time just after $k$-th pulse to that after $(k+1)$-th one. In consequence, we perform some sort of quantum mapping procedure and compare its numerical results with those from our analytical formulas. Fig.1a shows probabilities for four states ($|0\rangle_a |0\rangle_b, |0\rangle_a |1\rangle_b, |1\rangle_a |0 \rangle_b$ and $|1\rangle_a |1\rangle_b $) involved in the system's evolution. We see very good agreement between numerical (cross marks) and analytical results (lines). It should be stressed out that numerical results presented in this figure were obtained for basis involving considerably more than four states appearing in the definition of $|\Psi\rangle_{cut}$ (\ref{eq4}) -- we assumed $15$ states for each of two modes. Moreover, Fig.1b shows the deviation of the sum of the probabilities corresponding to our analytical result unity. We see that its amplitude is $\sim 10^{-3}$. In fact, this result shows how the fidelity between cut wave-function $|\Psi\rangle_{cut}$ and its "full" numerical counterpart $|\Psi\rangle$ differs from the unity. Results presented in Fig.1 indicate very good agreement between our analytical solution and results obtained from numerical simulations. It is seen that for the exemplary parameters assumed there our system behaves as \textit{nonlinear quantum scissors} \cite{LK11} -- we assumed that couplings are much smaller than nonlinearity constants. From other side, our system can be treated as \textit{qubit-qubit} one, because we have only two possibilities for each of the modes - vacuum state or one-photon state.
\begin{figure}
\begin{center}
\raisebox{5cm}{a)}\includegraphics[width=0.45\textwidth]{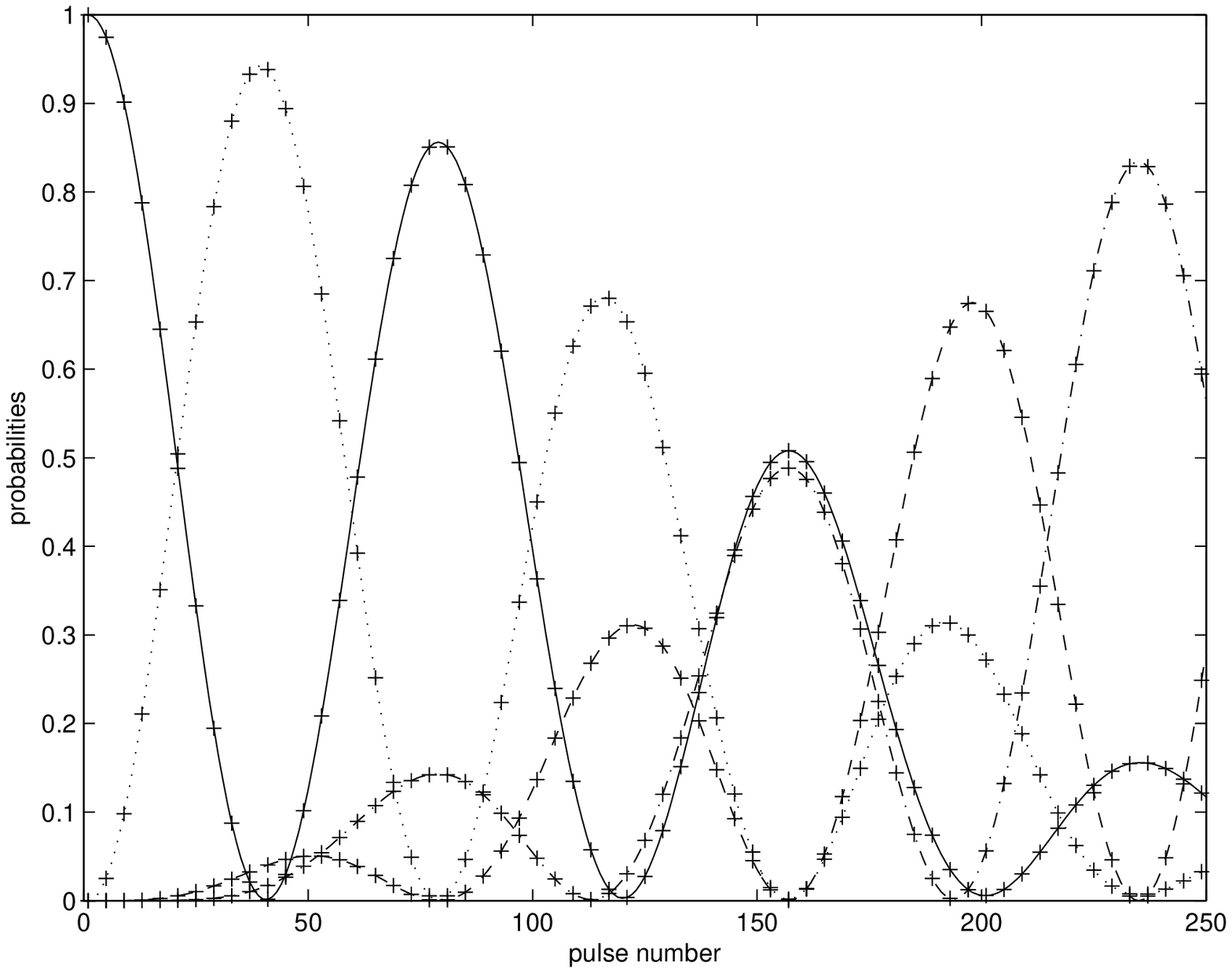}
\raisebox{5cm}{b)}\includegraphics[width=0.45\textwidth]{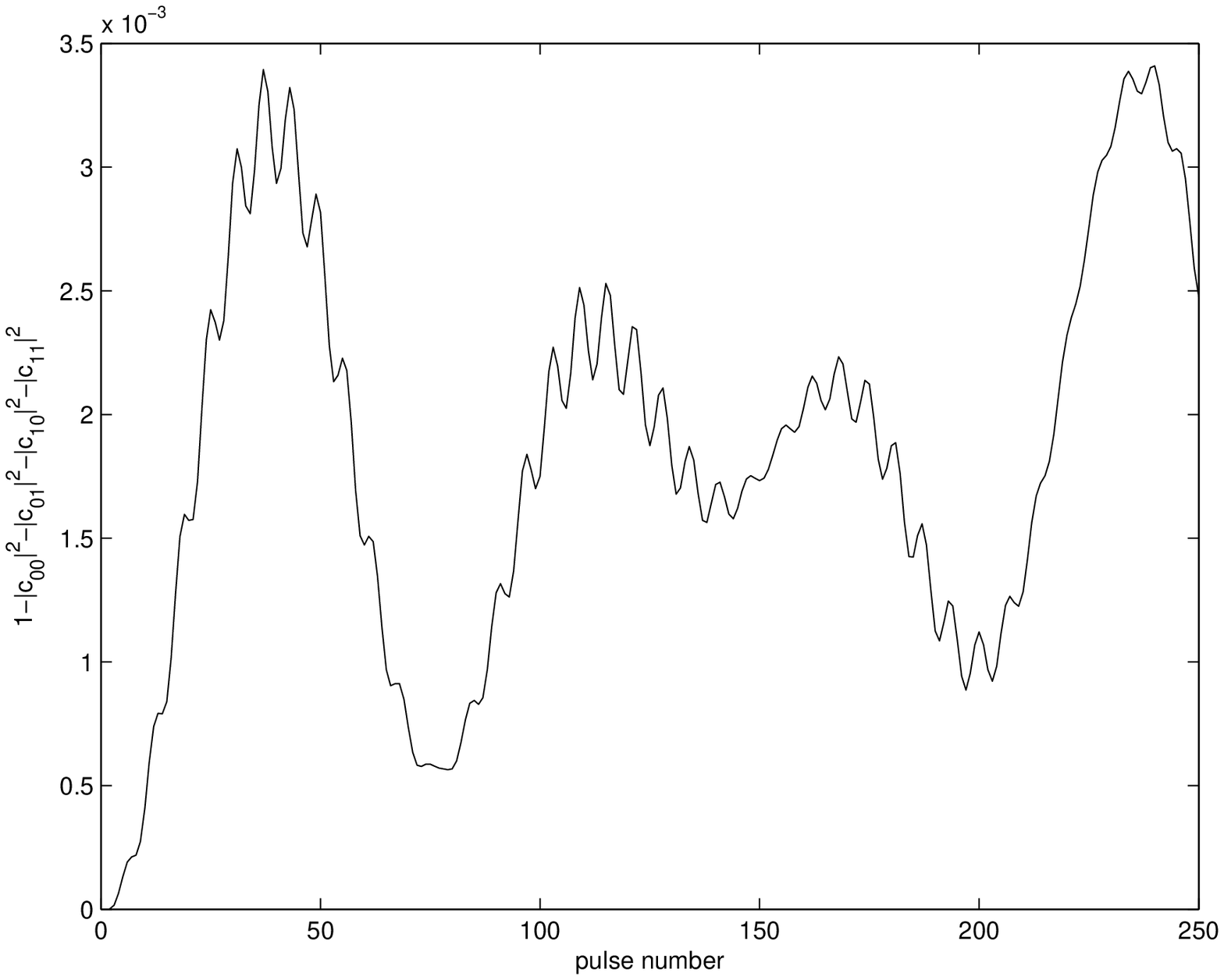}
\caption{The probabilities (a) for the states: $|0\rangle_a|0\rangle_b$ -- solid line, $|0\rangle_a|1\rangle_b$ -- dashed line, $|1\rangle_a|0\rangle_b$ -- dotted line, $|1\rangle_a|1\rangle_b$ -- dash-dotted line. Cross marks correspond to numerical results. We assume that $\alpha =1/25$, $\epsilon =1/100$ and $T=1$. All energies are expressed in units of nonlinearity constant $\chi_a=\chi_b=1$. The deviation  $1-|c_{00}|^2-|c_{01}|^2-|c_{10}|^2-|c_{11}|^2$ shown in (b) corresponds to the same parameters.}
\end{center}
\label{fig1}
\end{figure}

\section{Results and discussion}
It is seen from Fig.1a that for some moments of time the probabilities corresponding to the states $|0\rangle_a|0\rangle_b$ and $|1\rangle_a|1\rangle_b$ becomes simultaneously closed to $1/2$. Moreover, we can observe a similar situation for the pair $|0\rangle_a|1\rangle_b$ and $|1\rangle_a|0\rangle_b$, although for this case the values of the maxima of these probabilities differs from $1/2$. Therefore, we can expect that at least states closed to maximally entangled states (MES) could be generated in our model. Therefore, we calculated concurrence describing entanglement present in our system. It is defined as \cite{HW97,W98}
\begin{equation}
C(\hat{\rho})=max\{0, \lambda_1-\lambda_2-\lambda_3-\lambda_4\}
\label{eq8}
\end{equation}
where $\lambda_i$ are square roots of eigen-values, in decreasing order, of the matrix \linebreak
$\hat{\rho}(\hat{\sigma}_{y}^{a}\otimes\hat{\sigma}_{y}^{b})
\hat{\rho}^*(\hat{\sigma}_{y}^{a}\otimes\hat{\sigma}_{y}^{b})$ (operators $\hat{\sigma}_{y}^{a}$ i $\hat{\sigma}_{y}^{b}$ are Pauli matrices for the modes/qubits $a$ and $b$, respectively). Thus, Fig.2 shows how the concurrence changes with time for the same parameters as those for Fig.1.
\begin{figure}
\begin{center}
\includegraphics[width=0.4\textwidth]{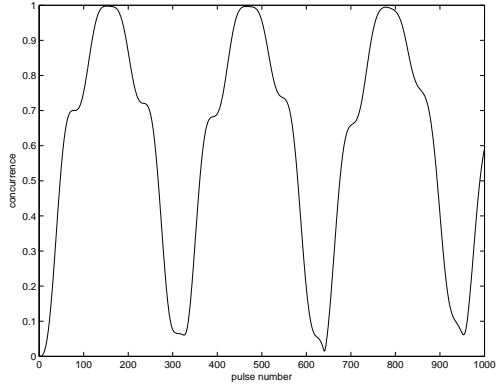} 
\end{center}
\caption{Time-evolution of concurrence. Parameters describing system are the same as those for Fig.1.}
\label{fig2}
\end{figure}
We see that it reaches its maximal values equal to $1$ repeatedly, so  we get MES. The moments of time corresponding to the generation of MES are the same as those for which $|c_{00}|^2\simeq |c_{11}|^2\simeq 1/2$. That means that at those moments of time Bell states are produced. Moreover, each maximum shown in Fig.2 is rather broad and additionally, is accompanied by two satellite maxima. This is an effect of the fact that when the probabilities $|c_{01}|^2$ and $|c_{10}|^2$ reach their own maximal values, $|c_{00}|^2$ and $|c_{11}|^2$ become close to zero.  For the moments of time when we observe such features other Bell states are generated as well, although with less accuracy. To check which Bell states appear in the system we calculate the fidelities between four Bell states and the wave function $|\Psi\rangle_{cut}$. The Bell states are:
\begin{eqnarray}
|B\rangle_1\,=\,\frac{1}{\sqrt{2}}\left( |0\rangle_a|0\rangle_b+
i|1\rangle_a|1\rangle_b\right),\quad
|B\rangle_2\,=\,\frac{1}{\sqrt{2}}\left( |0\rangle_a|0\rangle_b-
i|1\rangle_a|1\rangle_b\right),\nonumber \\
|B\rangle_3\,=\,\frac{1}{\sqrt{2}}\left( |0\rangle_a|1\rangle_b+
i|1\rangle_a|0\rangle_b\right),\quad
|B\rangle_3\,=\,\frac{1}{\sqrt{2}}\left( |0\rangle_a|1\rangle_b-
i|1\rangle_a|0\rangle_b\right).
\label{eq9
}\end{eqnarray}
\begin{figure}
\begin{center}
\raisebox{4.5cm}{a)}\includegraphics[width=0.4\textwidth]{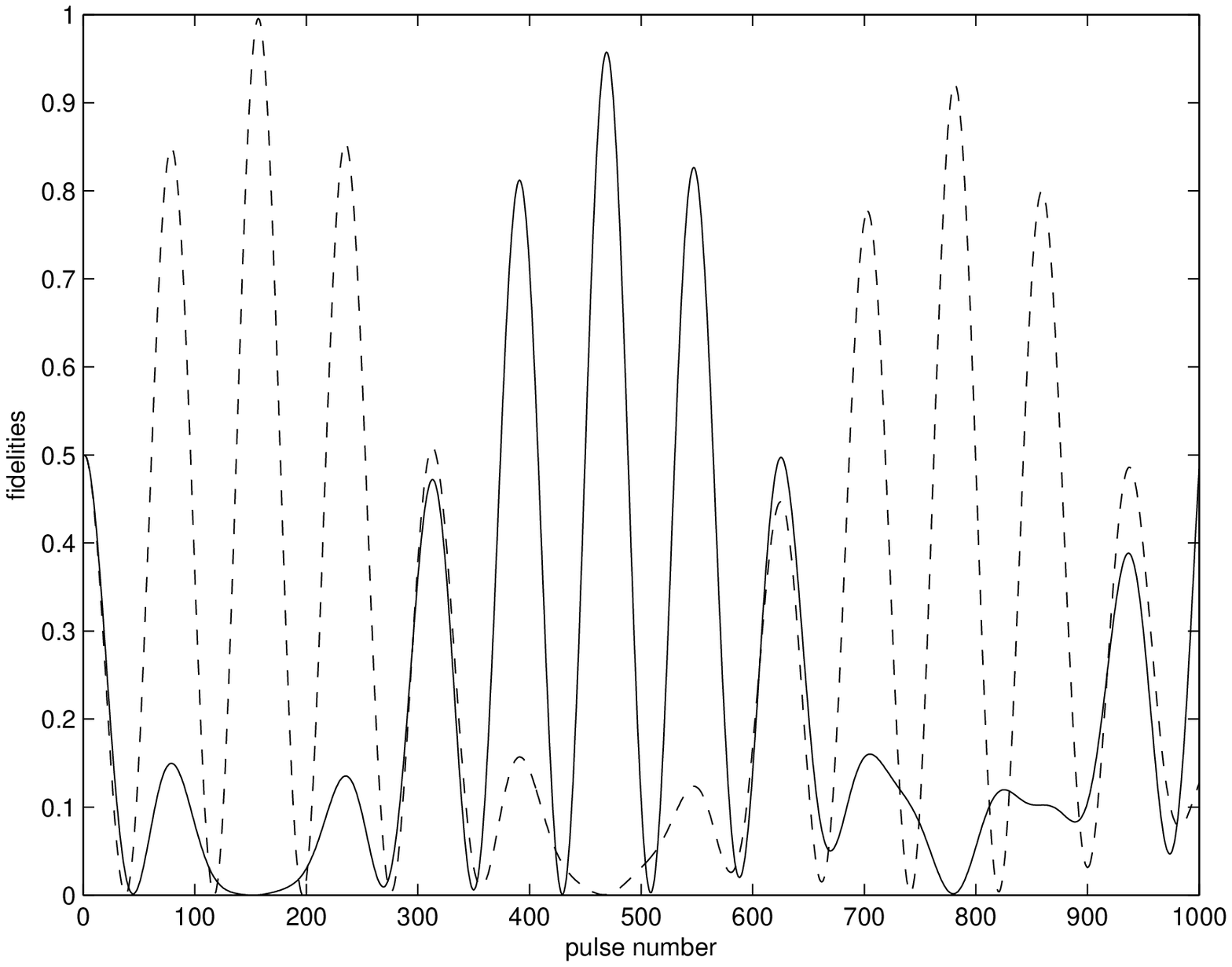} 
\raisebox{4.5cm}{b)}\includegraphics[width=0.4\textwidth]{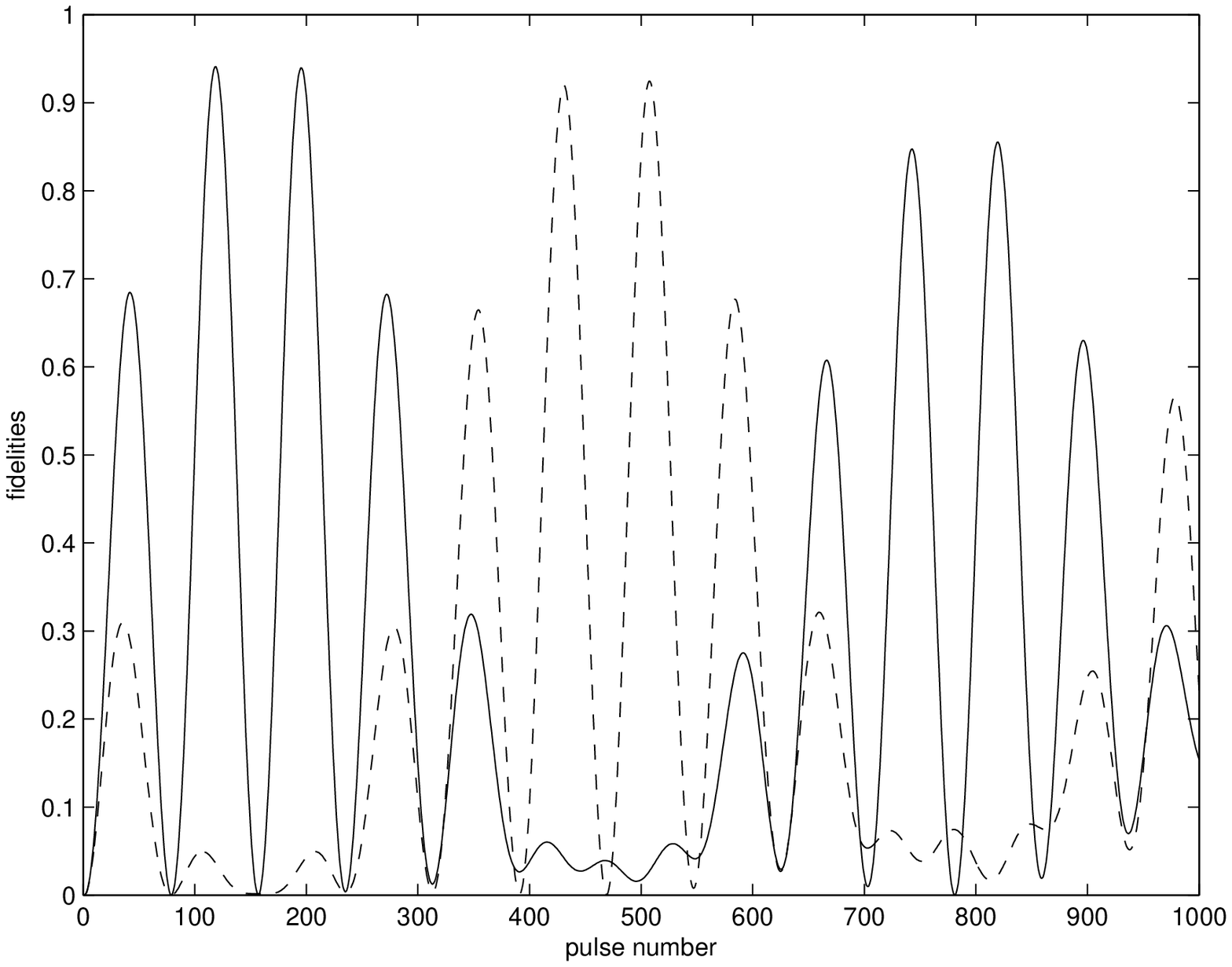} 
\end{center}
\caption{Fidelities corresponding to the Bell states: (a) $|B\rangle_1$ --dashed  line, $|B\rangle_2$ -- solid line, and (b) $|B\rangle_3$ -- dashed line, $|B\rangle_4$ -- solid line. The parameters are the same as for previous figures..}
\label{fig3}
\end{figure}
From Fig.3 we see that the state $|B\rangle_1$ can be generated almost perfectly (first maximum for Fig.1a, dashed line). Moreover, other mentioned here Bell states could be produced but with slightly less accuracy. The situation  resembles that discussed in \cite{LM04} but here we deal with the system excited by pulses instead of continuous external field with constant amplitude. For the case discussed here, we have additional parameter, time between two subsequent pulses $T$, that can be applied for tuning the system. Here the system evolution can be divided into two stages. First of them is evolution during extremely short pulses when interaction with external field plays crucial role and energy of the system is changed. The second stage is related to the system's ''free'' evolution during the period of time between two subsequent pulses. For that time the energy of the system is conserved and the phase factor related to the presence of nonlinearities is dominant. It is completely different mechanism than that presented in the system discussed in \cite{LM04} where these two factors, pumping and phase evolution, act simultaneously.

\section{Conclusions} 
We discussed a system of Kerr-like coupler excited by series of ultrashort pulses. We derived analytical formulas for the probability amplitudes and showed that the system can evolve as nonlinear quantum scissors and behaves as \textit{qubit-qubit} system. Moreover, it can be treated as a source of various Bell states. The model differs from that with continuous excitation where the solutions for probability amplitudes were different from those discussed here. It gives new potential possibilities of controlling  evolution of the model and engineering various quantum states of the field.

\section*{Acknowledgments}

A. K-K would like to thank NCN grant No DEC-2011/03/B/ST2/01903  for the support.

\end{document}